# A three-dimensional model for artificial atoms and molecules: Influence of substrate orientation and magnetic field dependence

**V. Mlinar* and F. M. Peeters**

*Departement Fysica,Universiteit Antwerpen, Groenenborgerlaan 171, B-2020 Antwerp, Belgium*

A full three–dimensional model for the calculation of the electronic structure of semiconductor quantum dots (QD) and molecules (QDM) grown on high index surfaces and/or in the presence of an external magnetic field is presented. The strain distribution of the dots is calculated using continuum elasticity and singe-particle states are extracted from the nonsymmetrized eight-band **k.p** theory. The model properly takes into account the effects of different substrate orientation by rotation of the coordinate system in the way that one coordinate coincides with the growth direction, whereas the effects of a tilted external magnetic field are taken into account throught the Zeeman effect and employing a gauge invariant scheme based on Wilson's formulation of lattice gauge theory. We point out the role of piezoelectricity for InAs/GaAs QDs grown on [11*k*], where *k* = 1,2,3,4,5,7,9 and for QDMs containing eight InAs/GaAs QDs grown on [11*l*], where *l* = 1,2,3. We predict the variation of the transition energies of the QDM as a function of substrate orientation and interdot distances in the molecule. We address the magnetic field direction dependent variation of the electronic properties of QD and QDM.

## I. Introduction

Semiconductor quantum dots (QDs) are three-dimensional (3D), artificial, semiconductor based structures with size of a few nanometars in all three directions.[1,2] Their confinement, smaller than de Broglie wavelength in semiconductors, leads to a discrete energy spectrum and a delta–function atomic–like density of states, which enables the analogy with real atoms, and therefore, QDs are ofter referred to as artificial atoms. Furthermore, the coupling between QDs to obtain new functional units,[3] leads to quantum dot molecules (QDM), and extends the analogy between real and artificial atoms to the molecular realm.[4]

The growth conditions mainly determine the properties of QDs and QDMs.[5] Stranski-Krastanow (SK) growth mode among all other methods is currently widely used for self-assemble QD formation and is driven by local strain fields that enables the formation of coherent islands.[1,2,5] Also, the same growth mode enables the formation of vertically stacked QDM, driven by the strain that surrounds a quantum dot in the lower lying layer which enforces the location of the next QD on top of the first one.[6,7] The occurrence of strain fields in and around the dots and significant intermixing, tightly connected with the SK growth mode, as well as multi-band and inter-valley mixing present in the QD systems leads to new features in QDs and QDMs as compared to real atoms and molecules (e.g. see Ref. 8).

Manipulation of the QDs' properties is driven by current and potential applications of QDs and QDMs, ranging from QD lasers,[9] a possible physical representation of a quantum bit,[10] to nanosensors used to detect DNA molecules.[11] Growth of QDs on high index planes is one of the ways to control the QD properites. From the experimetal point of view, QDs' growth on high index surfaes may lead to an increased QD density and a lower size dispersion as compared to the well–known [001] grown QDs. It also enables 3D growth ranging from a chainlike pattern to a squarelike lattice of QDs.[12] From a physics point of view, it may result in the variation of the photoluminescence energy with substrate orientation. This is due to the different planar projections of conduction and valence bands of the constituent crystals forming the QDs originating from different substrate orientations.

Furthermore, an external magnetic field is an interesting tuning parameter that opens up a new playground for the investigation of spin-related properties.[13] It also leads to a competition between quantum confinement, Coulomb interaction and magnetic field confinement.[14]

A thorough theoretical analysis should include a realistic geometry and composition profile of the QD, and take into account the strain distribution in and around the dots, effects of piezoelectricity, and multi-band mixing. The electronic structure calculations of QD and QDMs are usually performed using **k.p**[15-23] and atomistic methods,[24-27] where inter-valley mixing is included as well. Furthermore, most of the previous theoretical work was restricted to [001] grown QDs.

In this paper, we present the 3D nonsymmetrized eight–band **k.p** model for electronic structure calculation of QDs and QDMs grown on high index planes and/or in the presence of an external magnetic field. Our model properly takes into account the effects of the different substrate orientation by rotation of the coordinate system in the way that one coordinate coincides with the growth direction. The effects of a tilted external magnetic field are taken into account throught the Zeeman effect and we employ a gauge invariant grid discretization based on Wilson's formulation of the lattice gauge theory. In the framework of the presented model, we analyze the effects due to substrate orientation on the piezoelectricity and the transition energies for InAs/GaAs [11*k*] grown QDs, where $k$ = 1, 2, 3, 4, 5, 7, 9, and InAs/GaAs [11*l*] QDM consisting of eight QDs, where $l$ = 1, 2, 3, and compare them to the ones of the well investigated [001] grown QDs and QDMs, respectively. Influence of the direction of an applied magnetic field on the electron and hole states of a single lens shaped QD, and QDM is further addressed.

* E-mail: vladan.mlinar@ua.ac.be; francois.peeters@ua.ac.be

The paper is organized as follows. Our theoretical approach is presented in Sec. II. The details of our numerical implementation are given in Sec. III. Sec. IVA contains the numerical results and discussion of the effect of substrate orientation on the piezoelectricity and the transition energies. The effects of an external magnetic field on the electronic properties of QDs and QDM are discussed in Sec. IVB. Our results and conclusions are summarized in Sec. V.

## II. Choice of the model

### A. Theory versus experiment

We define now the different steps in the modelling of the QDs' electronic and optical properties and specify at which point the experimetal data are linked to our model.

(i) Size, shape and composition profile of our model QD and/or QDM as they enter into our calculations are extracted from structural characterization, e.g. by cross section scanning tunneling microscopy or HRTEM.[1,2,5] It is also important to note that the strain and piezoelectric fields in and around the dots and the compositional intermixing during the QD growth lead to uncertainties in the determination of their geometry and composition and therefore limits the predictive power of the theoretical models. (ii) In the second step we calculate the strain distribution in and around the dots using continuum elasticity theory.[28-30] The character of the strain is determined by the hydrostatic *Tr(e)* part of the strain tensor influencing the conduction and valence band profiles and the biaxial part *B(e)* of the strain tensor influencing the valence band mixing.

**Table 1** Material parameters used for the electronic structure calculations

| Num. | Parameter | Units | GaAs | InAs |
|---|---|---|---|---|
| *1* | $a_{latt}$ | *Å* | 5.6503 | 6.0553 |
| *2* | $a_c$ | *eV* | -7.17 | -5.08 |
| *3* | $a_v$ | *eV* | 1.16 | 0.66 |
| *4* | $b$ | *eV* | -2.0 | -1.8 |
| *5* | $d$ | *eV* | -5.062 | -3.6 |
| *6* | $C_{11}$ | GPa | 121.1 | 83.3 |
| *7* | $C_{12}$ | GPa | 54.8 | 45.3 |
| *8* | $C_{44}$ | GPa | 60.4 | 39.6 |
| *9* | $e_{14}$ | C/m$^2$ | 0.16 | 0.045 |
| *10* | $\varepsilon_r$ | | 13.18 | 14.6 |
| *11* | $E_0$ | *eV* | 1.52 | 0.42 |
| *12* | $Ev$ | *eV* | -5.622 | -5.449 |
| *13* | $E_P$ | *eV* | 25.7 | 22.2 |
| *14* | $D_{SO}$ | *eV* | 0.34 | 0.38 |
| *15* | $m_{eff}$ | $m_0$ | 0.067 | 0.023 |
| *16* | $\gamma_1$ | | 7.65 | 19.67 |
| *17* | $\gamma_2$ | | 2.41 | 8.37 |
| *18* | $\gamma_3$ | | 3.28 | 9.29 |
| *19* | $\kappa$ | | 1.72 | 7.68 |

[a] Parameters for $In_xGa_{1-x}As$ are obtained as a linear interpolation between InAs and GaAs, except for the Luttinger parameters $\gamma_j$ where a harmonic average is performed (Ref. 15).

The off-diagonal components of the strain tensor lead to a piezoelectric potential which as a consequence influences the distribution of the electron and hole wave-function inside the dot. (iii) Using diagonalization of the eight-band Hamiltonian, including the strain and piezoelectric potential, the confined electron and hole energy levels are obtained as a next step in our modelling. Note that **k.p** theory is a semi – empirical theory, and is parametrized by a set of experimentally obtained twelve parameters at the $\Gamma$ point (see Table I, parameters *2–5* and *11–18*). For the presnet case of nanostructures, for each of the materials forming the nanostructure, the values of the paramteres are taken from the corresponding bulk material. At this point it is important to stress that no adjustable paramters are present in this model. (iv) As a final step in our model, we calculate the energy of the exciton complexes and the absorption spectra. It is at this point where we compare our findings with the experimental data and two cases can be distinguished. First, if the available photoluminescence measurements are performed on an ensemble of QDs, the information about the individual characteristics of the QDs are lost, and therefore we consider our model QD as a representative of the ensemble of QDs. Second, if the experimetalists provide us with single dot spectroscopy data, our model QD has to include the exact geometry and composition profile as the ones of the dot which the experiments were performed on. Only in that case one can expect that the measured optical spectra is correctly interpreted. In this work we will focus on the implementations of the points (i)– (iii) including direct Coulomb interaction.

### B. Conventional versus nonsymmetrized Hamiltonian

The bandstructure of the semiconductors in the presence of weak perturbing inhomogeneous potentials is sucessfully described in the framework of the multiband effective-mass theory.[31,32] However, the validity of this theory is questionable when applied to nanostructures.[33,34] The most common implementation is based on the usage of the appropriate bulk Hamiltonian for each constituent material separately, and the envelope functions on either side of the interface are connected by applying *ad hoc* symmetrization rules.[35] This approach is often referred to as the conventional approach, and the corresponding Hamiltonain as the conventional Hamiltonian. We have shown, in our earlier publication,[36] that such implementation leads to nonphysical solutions in case of nanostructures having a large difference of the effective-mass parameters between the constituent materials. Our proposed improvement is an approach based on the envelope-function theory for nanostructures developed by Burt and Foreman[33,34] and is referred to as the nonsymmetrized approach and the corresponding Hamiltonian as the nonsymmetrized Hamiltonian. The term "nonsymmetrized" is coming from the fact that although the Hamiltonian has to be Hermitian, its matrix elements may not be.[34] In case of nanostructures the $\boldsymbol{k}$ operators fail to commute with the effective-mass parameteres at the heterointerface of the nanostructure. In a non-zero magnetic field the $\boldsymbol{k}$ operators also fail to commute between themselves. Actually, the additional terms of the nonsymmetrized Hamiltonian which appear because of the variation of the effective-mass parameters at the interface are equivalent to the Zeeman energy terms appearing in the Hamiltonian due to the magnetic field. Although several subtile effects were not included in the nonsymmetrized model,[37] it has been shown that its implementation does not lead to non-physical solutions.[36]

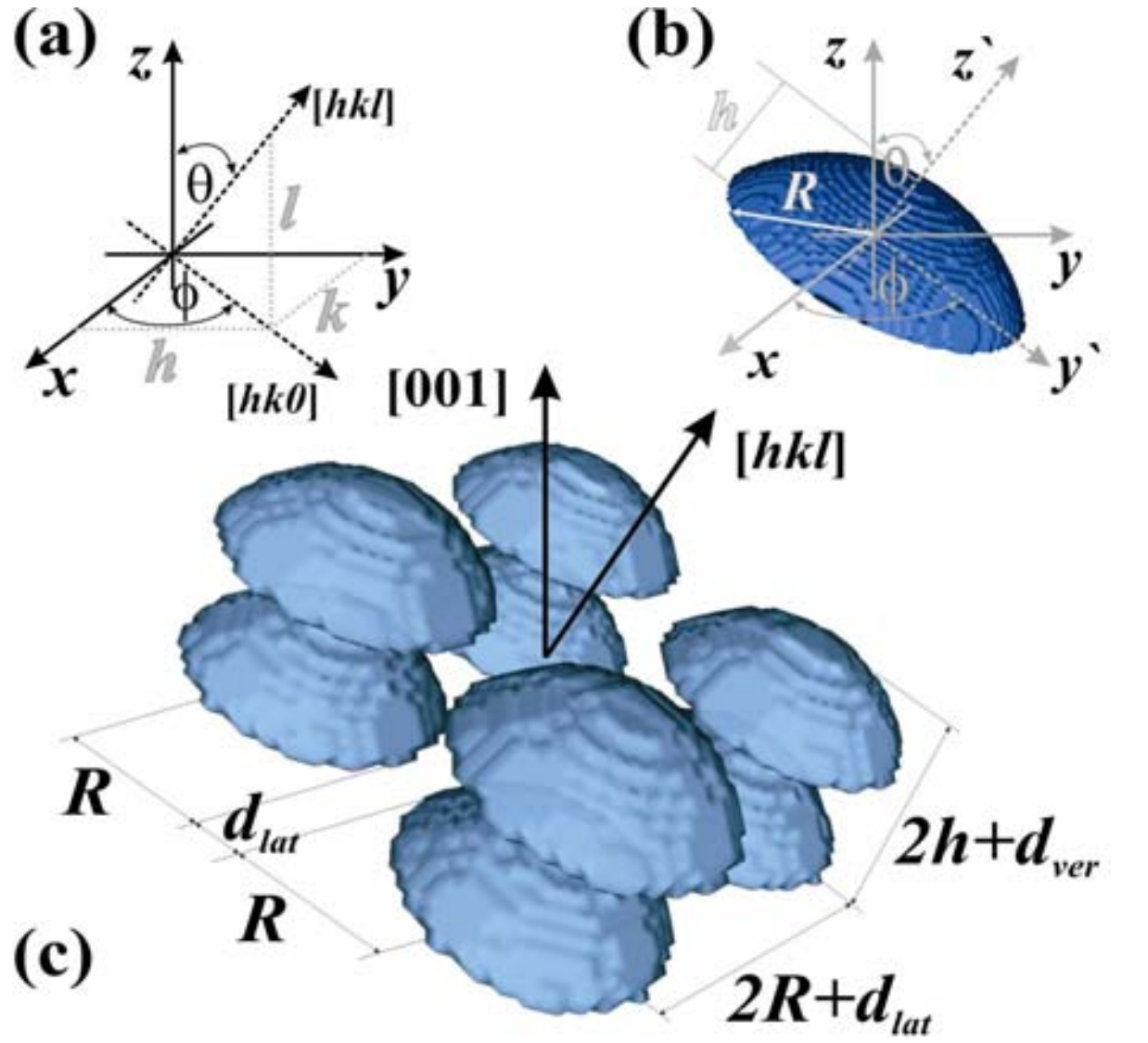

**Fig. 1 (Colour online)** (a) General coordinate system (x`,y`,z`) related to the [001] system (x,y,z) through a transformation matrix *U(ϕ, θ)*, where *ϕ, θ* are the azimuthal and polar angle, respectively. (b) Model lens-shaped [*hkl*] grown QD of radius *R* and height *h*. (c) Model quantum dot molecule formed from eight identical lens-shaped QDs of radius R and height h, where $d_{lat}$ is the inter-dot distance in the lateral direction, and $d_{ver}$ is the inter-dot distance in the growth direction.

### C. Modeling of quantum dots grown on high-index surfaces

QDs' growth on high index (*hkl*) surfaces can be included in our model in two ways. First, one can keep the employed Hamiltonian related to the [001] coordinate system and define the structure in the [*hkl*] direction. Although the optical properties can be extracted from such a model (see e.g. Ref. 38), the underlying physical features of the QD system cannot be traced. The second approach is based on the rotation of the coordinate system and the employed Hamiltonian, in the way that the Carthesian coordinate *z'* coincides with the growth direction as shown in Fig. 1.

The general [*hkl*] coordinate system (x`, y`, z`) is then related to the conventional [001] coordinate system (x, y, z) through the transformation matrix $U = U(\phi, \theta)$ [39,40]

$$U = \begin{pmatrix} \cos\phi\cos\theta & \sin\phi\cos\theta & -\sin\theta \\ -\sin\phi & \cos\phi & 0 \\ \cos\phi\sin\theta & \sin\phi\sin\theta & \cos\theta \end{pmatrix}, \qquad (1)$$

The angles $\phi$ and $\theta$ represent the azimuthal and polar angles, respectively, of the [*hkl*] direction relative to the [001] coordinate system. The azimuthal angle $\phi$ and the polar angle $\theta$ are specified in terms of indices *h*, *k*, and *l* and are given by:

$$\tan\phi = \frac{k}{h}\ , \quad \tan\theta = \frac{\sqrt{h^2+k^2}}{l}. \qquad (2)$$

The strain components related to the rotated coordinate system, are denoted by $\varepsilon'_{ij}$, where *i, j = x, y, z*, and are derived by applying the unitary transformation to the stiffness coefficient $C_{ij}$ tensors according to the relationship:

$$C'_{ijkl} = U_{im}U_{jn}U_{ko}U_{lp}C_{mnop}. \qquad (3)$$

Furthermore, the relation between $\varepsilon'_{ij}$ and $\varepsilon_{ij}$, where $\varepsilon_{ij}$ are the strain components related to the [001] coordinate system, is given by:

$$\varepsilon_{ij} = U_{ki}U_{lj}\varepsilon'_{kl}. \qquad (4)$$

The shear strains induce a piezoelectric polarization which creates fixed charges and piezoelectric potentials. In the case of QDs and QDMs grown on high index planes the magnitude of the polarization field is proportional to the off-diagonal components of the strain tensor related to the [001] coordinate system.

The appropriate Hamiltonian for the [*hkl*] crystallographic orientation is derived by rotating the angular and crystallographyc orientation in accordance to the relationship:[39-42]

$$k'_i = U_{ij}k_j\ , \qquad (5)$$

where $i, j = x, y, z$.

We applied here the Hamiltonian, as given in Ref. 36, to our QD structures. We stress that, by nature of its derivation, the nonsymmetrized Hamiltonian is valid for any growth direction, and one can simply replace the old ***k*** operators by their projections in the new frame (as given above).[43]

### D. An external magnetic field

In the framework of the multi-band effective-mass theory, a magnetic field is included through Peierls substitution in the wave-vector and by adding the Zeeman energy term.[31,32,44] However, it has been proven that Peierls substitution in the Hamiltonian, which subsequently is discretized on a grid, may lead to the breaking of gauge invariance.[45] As a result of such a straightforward implementation, large errors in the estimates of the eigenvalues and corresponding eigenvectors may occur, producing solutions that do not reflect the actual physics, but rather ill-employed numerics. Therefore, a discretization scheme which preserves gauge invariance is required. Here, we implement a gauge invariant grid discretization based on Wilson's formulation of the lattice gauge theory.[46]

For the discretized Hamiltonian we have to impose the following condition: The discretized Hamiltonian has to maintain all its physical properties if we perform a gauge transformation of the vector potential.[46] Or more formally: If the vector potential transforms as $\vec{A} \to \vec{A} + \nabla\chi$, then a wave function representing a physical state of the system transforms as $\psi(\vec{r}) \to G(\vec{r})\psi(\vec{r})$, where $G(\vec{r}) = \exp(-i(e/\hbar)\chi(\vec{r}))$ (*e* – electron charge, $\hbar$ - Planck constant). As a consequence, the Hamiltonian transforms as $H \to G(\vec{r}) H G^{+}(\vec{r})$. Furthermore, considering a uniform discretization grid, the lattice operator $U_j(\xi)$ can be defined as $U_j(\xi)=exp(-i(e/\hbar)a_{latt}A_j(\xi))$, where $j = x, y, z$, and $a_{latt}$ is the grid spacing, and $\xi = (la_{latt}, ma_{latt}, na_{latt})$ defines the positon on the grid. The lattice operator transformes as: $U_j(\xi) \to G^{+}(\xi) U_j(\xi) G(\xi + j)$. Note that the $U_j(\xi)$ can also be understood as the link variable between two points on the grid. Previous relations lead to the correct discretization scheme, and as an example, the discretization of the second derivative is given by Eq. (6).

$$\frac{\partial^2 \psi}{\partial x^2} \to \frac{U_x^{+}(i,j,k)\psi_{i+1,j,k} - 2\psi_{i,j,k} + U_x(i+1,j,k)\psi_{i-1,j,k}}{a_{latt}^2}. \tag{6}$$

We stress that the correct choice of the boundary conditions at the edges of the calculation box is very important. While the phase of the wave function is preserved on the inner points of the grid, as shown above, the phase should also be preserved at the boundary. Imposing Dirichlet boundary conditions ensures phase preserving at the edges of the computational box.

For completeness, let us add that Wilson's formulation, as presented here, is oriented more to the numerical implementation, and does not explicitly exploit the periodicity of the crystal potential. In contrast, a general theory for the nonperturbative Bloch solution of Schrödinger's equations in the presence of a constant magnetic field was recently proposed,[47] where an equivalent quantum system with a periodic vector potential was obtained using a single gauge transformation based on a lattice of magnetic flux lines.

## III. Numerical implementation of the 3D k.p model

The experimental uncertainties in the geometry and composition profile of a QD or QDM enforce us the usage of numerical procedures capable for easy handling of variations of 3D input structures. Therefore, we will create our model structure on a 3D rectangular grid which enables us to define parameters of our model at each node of the grid.

Strain is determined by global minimization of the action integral for elastic strain.[28-30] The model takes into account the finite size of the dots and includes the anisotropy of the elastic modulus. The components of the strain tensor are calculated directly from the displacement *u* that minimizes the action integral for the strain tensor.[29,30] The displacements are discretized at the nodes of the grid representig their first derivatives by finite differences. The first derivative is averaged over the eight permutations of forward and backward differences.[29,30] Typically, the grid consists of N = 189×189×159 nodes. Numerically, a matrix equation Au =b is solved by the conjugate gradient method (CGM), where *A* is a 3N×3N matrix, and *b* is a known vector of size 3N. The piezoelectric potential is extracted from the Poisson equation, taking into account the image charge effects due to the discontinuous dielectric constants at the hetero-interfaces. Although these image charge effects are not significant for our embedded QDs, they become very important in the case of uncapped QDs. The Poisson equation is discretized on a larger grid than the one used for the strain calculations, we took tipically N=199×199×199 and is solved by CGM.

For the calcuation of confined electron and hole states, we significantly reduce the size of our 3D grid in all three directions. The reason lies in the fact that while strain decays slowly, with a power law, the electron and hole wave functions have an exponential decay. Therefore, our grid for the electronic structure calculation is typically N=69×69×59. In the case of the QDMs contaning eight dots, as will be shown later, a larger grid of N=79×79×79 was used. Eight coupled Schrödinger equations including the strain Hamiltonian and the piezoelectric potential are discretized by the finite difference method, whereas special care is taken for the boundary nodes between the dot and the surrounding material.[48] The appropriate effective-mass parameters from Ref. 49 (see Table. I), and grid spacing are choosen such that no numerically caused spurious solutions appear.[50] In our calculation we use a grid spacing equal to the lattice constant of the material that surrounds the dots. Therefore, in the numerical experiment presented in next Section, the grid spacing equals $a_{latt}(GaAs)$ (see Table I). The finite-difference scheme is tested by checking the consistency and stability.[51] In order to fulfil the condition of stability, the problematic operators for discretization are: $\partial/\partial i$, $\partial^2/\partial i \partial j$ where $i, j = x, y, z$, and $i \neq j$. The discrete approximation to the first derivatives, from the physical point of view, should exhibit the same spectral characteristic, i.e. it should be non–dissipative and non–dispersive.[52] Non-dissipative approximation is a necessary condition for numerical stability to be independent of the propagation direction.

Confined electron and hole states are found by diagonalization of the discretized Hamiltonian, with dimensions of several million by several million. The Hamiltonian is diagonalized twice, once to obtain the confined electron states and a second time for the confined hole states. In this work the numerical implementation[53] of the Jacobi–Davidson (JD) algorithm[54] is used for extracting eigenvalues and eigenvectors of the matrix.

From the eigensolver we expect that it will give us a few (up to 50) inner eigenvalues and eigenvectors around the given searching energy. The JD algorithm is actually a combination of the Jacobi's orthogonal component correction (JOCC) approach and the Davidson method.[54] Beside the actual eigensolver loop, JD involves the solution of the linear system, i. e. the inner loop of the solver where the correction equatuon is solved. In the present implementation the Quasi Minimal Residual Simplified with preconditioning is used as the linear system solver. An overall acceleration of the nested loop (linear system / eigenvector) is achieved if, and only if, the JOCC preconditioner succeeds to increase the spectral gap of the desired eigenvalue so much that the entire additional effort spent for calculating the sophisticated correction vector is more than compensated. Finding the highly efficient, easy to calculate, preconditioner is a very tedious task and it is determined by the structure of the matrix.

# IV. Numerical results and discussion

## A. Growth on high index surfaces

It has been shown that growth of QDs on high index planes has practical advantages ranging from good quality QDs, in the sense of low size dispersion, to the possibility to control planar and vertical ordering of QDs lattices.[12] Here, we show the influence of substrate orientation on the character of strain, the role of piezoelectricity, and the transition energies of QDs and QDMs and compare it to the case of the well investigated [001] grown QDs and QDMs. Guided by the experimental findings of Ref. 12, we construct a QDM consisting of eight laterally and vertically coupled QDs, as shown in Fig. 1(c), as the first step in the investigation of the electronic and optical properties of QD lattices, and their dependence on the substrate orientation and inter-dot distances.

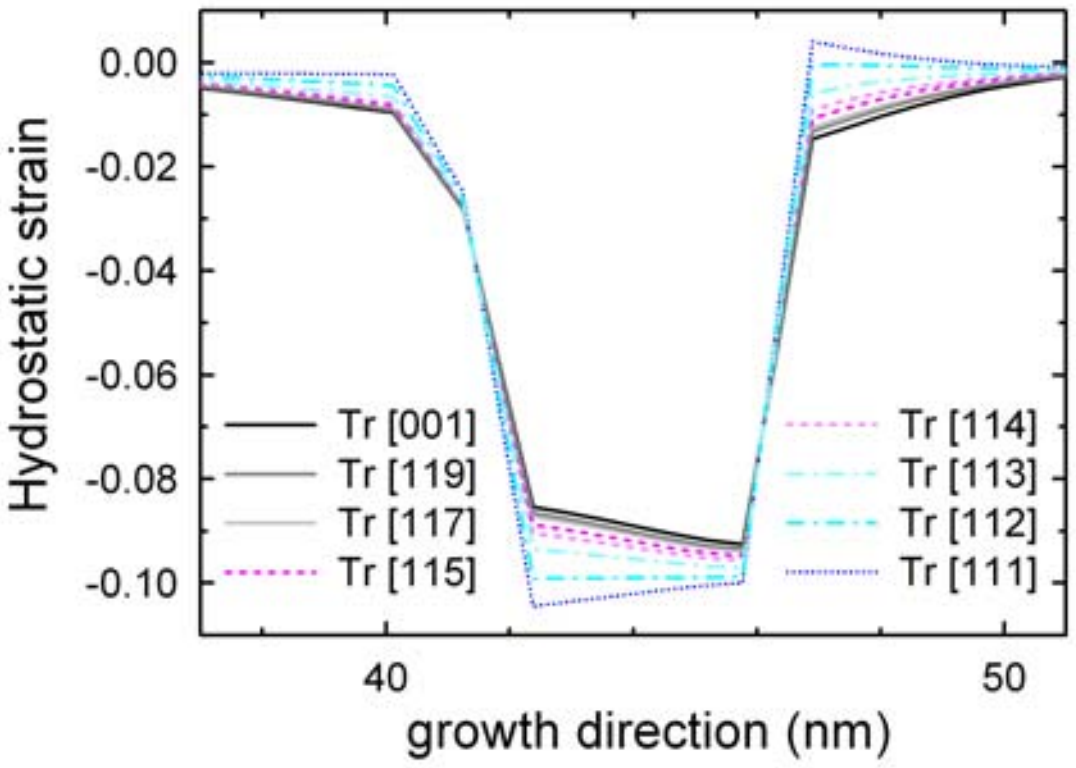

**Fig. 2 (Colour online)** Hydrostatic part of the strain tensor for lens shaped QD along the growth direction through the middle of the lens-shaped QD.

For numerical purposes we consider a lens shaped QD with radius R = 12.5nm and height h = 4.52nm (see Fig. 1(b)), and wetting layer of thickness 2.28nm and *In* concentration of 20%. Our model QDM consists of eight identical lens shaped QDs, as shown in Fig. 1(c), with radius R = 7.91nm, and height h = 4.52nm. We consider four different molecules, classified by the distance between the dots in the lateral direction $d_{lat}$ and the distance between the dots in the vertical direction $d_{ver}$ as given in Table 2. In the model of the QDM the wetting layer was not included.

**Table 2** QDM classified by the inter-dot distances in lateral, $d_{lat}$, and vertical, $d_{ver}$, direction.

| Molecule | $d_{lat}$ (nm) | $d_{ver}$ (nm) |
|---|---|---|
| M1 | 2.26 | 2.82 |
| M2 | 5.65 | 2.82 |
| M3 | 2.26 | 5.09 |
| M4 | 5.65 | 5.09 |

The hydrostatic component of the strain tensor along the growth direction is shown in Fig. 2 for our lens shaped model QD. The largest increase of the hydrostatic strain component is found for [111] grown QDs, whereas the [117] and [119] grown QDs exhibit almost negligeable difference with [001] grown QDs. Also, for [111] grown QDs the hydrostatic part of the strain tensor tends rapidly to zero in the barrier, which will be important for the QDM case, as discussed below.

**Table 3** Maximal and minimal value of piezoelectric potential for lens-shaped quantum dots as a function of the substrate orientation

| [*hkl*] | Max($V_{piezo}$) (mV) | Min($V_{piezo}$) (mV) |
|---|---|---|
| [001] | 53.10 | -53.10 |
| [111] | 75.99 | -70.67 |
| [112] | 66.12 | -68.79 |
| [113] | 59.34 | -64.62 |
| [114] | 55.74 | -61.51 |
| [115] | 53.67 | -59.43 |
| [117] | 52.52 | -56.93 |
| [119] | 51.85 | -55.42 |

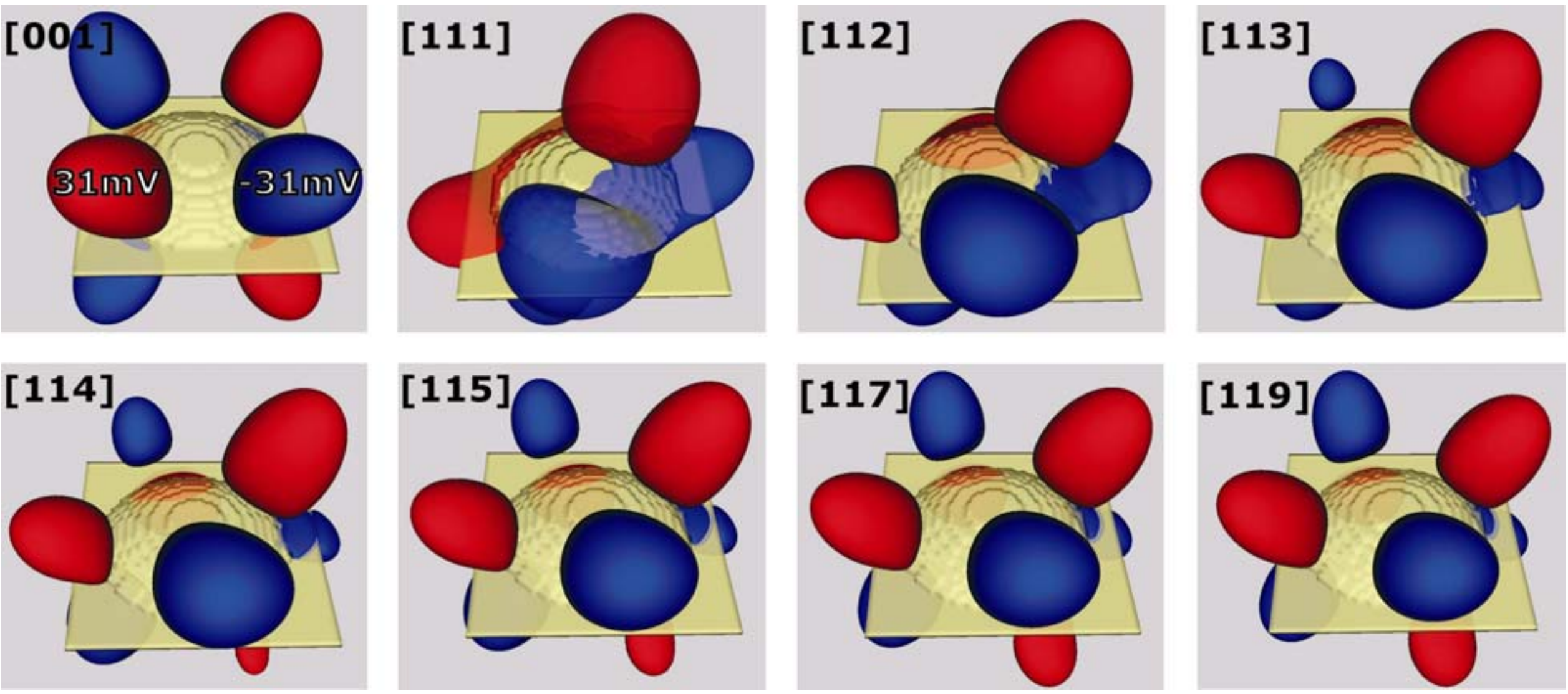

**Fig. 3 (Colour online)** Piezoelectric potential for a QD with isosurfaces at ±31mV for different substrate orientation.

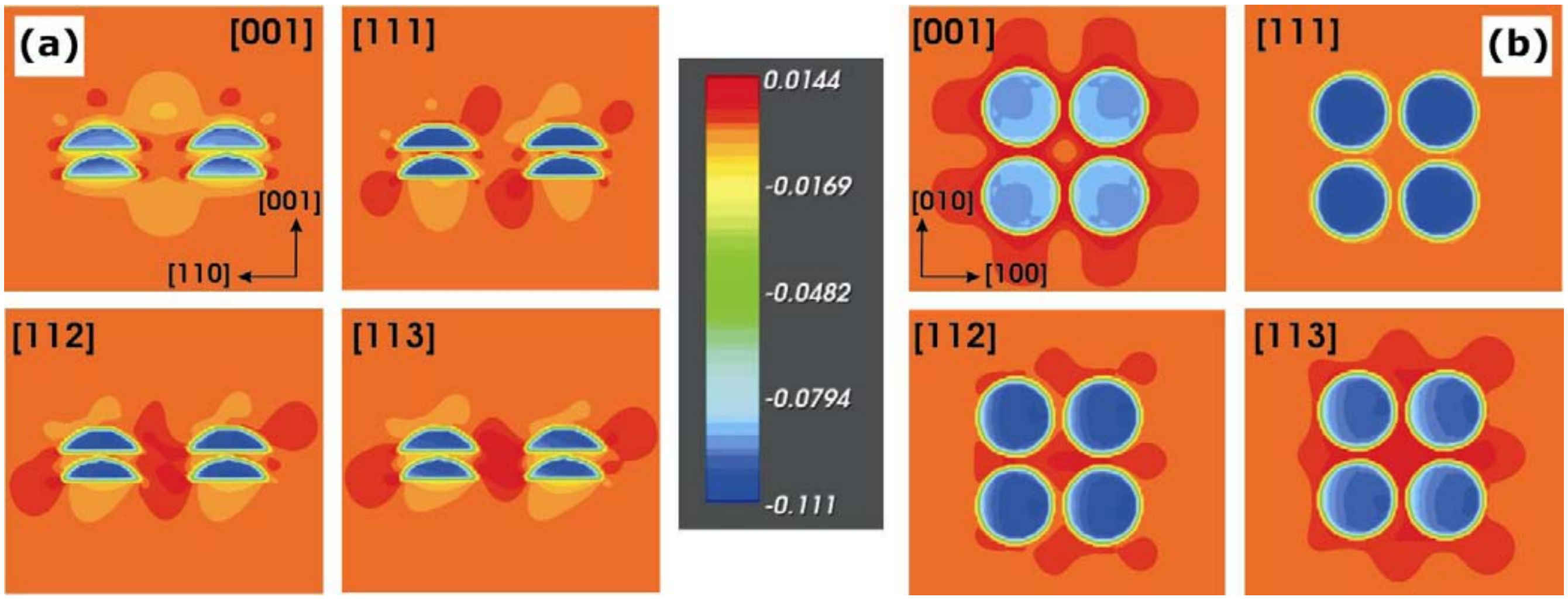

**Fig. 4 (Colour online)** Contour plot of the hydrostatic strain for M1 QDM (a) in a plane along the growth direction and the diagonal of the lateral dots in the molecule, and (b) in the plane perpendicular to the growth direction, at the bottom of the second layer of the QDs.

Discussion about the influence of the QDs' size and shape on the electronic and optical properties of [11*k*] grown QDs, where *k* = 1,2,3, was given in our earlier publication.[17]

Fig. 3 shows the piezoelectric potential of [11*k*] grown QDs, where *k* = 1, 2, 3, 4, 5, 7, 9, whereas the minimal and maximal values of the piezoelectric potential are given in Table 3. The piezoelectric potential of [001] grown QDs is also shown in Fig. 3 for comparative reasons. The largest piezoelectric potential is obtained for [111] grown QDs, reflecting also the three fold symmetry of the [111] direction. Going from the [111] grown QDs to [119] the value of the piezoelectric potential is lowered, and for [119] grown QDs, it approaches the case of our reference case of [001] grown QD (although asymmetric as compared to the reference case). Note that it was recently argued that the second order piezoelectric term, usually not included in the model, is of the same order of magnitude as the linear term, and is responsible for the lowering of the piezoelectric potential.[55] However, a different derivation of the second order piezoelectric term was also reported very recently,[56] showing a smaller contribution of the second order piezoelectricity to the total piezoelectricity effect as compared to the one proposed in Ref. 56. Therefore, in this work, we included only the linear term as is widely accepted in the literature.[28,15,22,26]

Hydrostatic strain profiles of the M1 QDM are shown in Fig. 4. Fig. 4(a) shows a cross section of the hydrostatic strain profile along the growth direction and along the diagonal through the lateral dots in the molecule. The hydrostatic strain profile in the plane perpendicular to the growth direction, at the bottom of the second layer of QDs is shown in Fig. 4(b). The largest strain coupling between the dots is observed for [113] grown QDM, whereas the strain coupling between the dots for [111] grown QDM is significantly reduced. This can be understood by looking at the hydrostatic strain for a single QD (see Fig. 2). For [111] the hydrostatic component is almost completely localized inside the dot. Only the case of the M1 molecule is shown since with increasing distance between the dots the coupling reduces. For QDMs, the role of piezoelectricity is increased as compared to the single dot case. Fig. 5 shows the piezoelectric profile of the [11*k*] grown QDM, where *k*=1,2,3. The reference case of the [001] grown molecule is shown as well. The very complex piezoelectric potential is a consequence of the fact that the dots forming the molecule are very close to each other (~2nm). Increasing the distance between the dots leads to piezoelectric potential profiles that are more similar to the single dot case (not shown here).

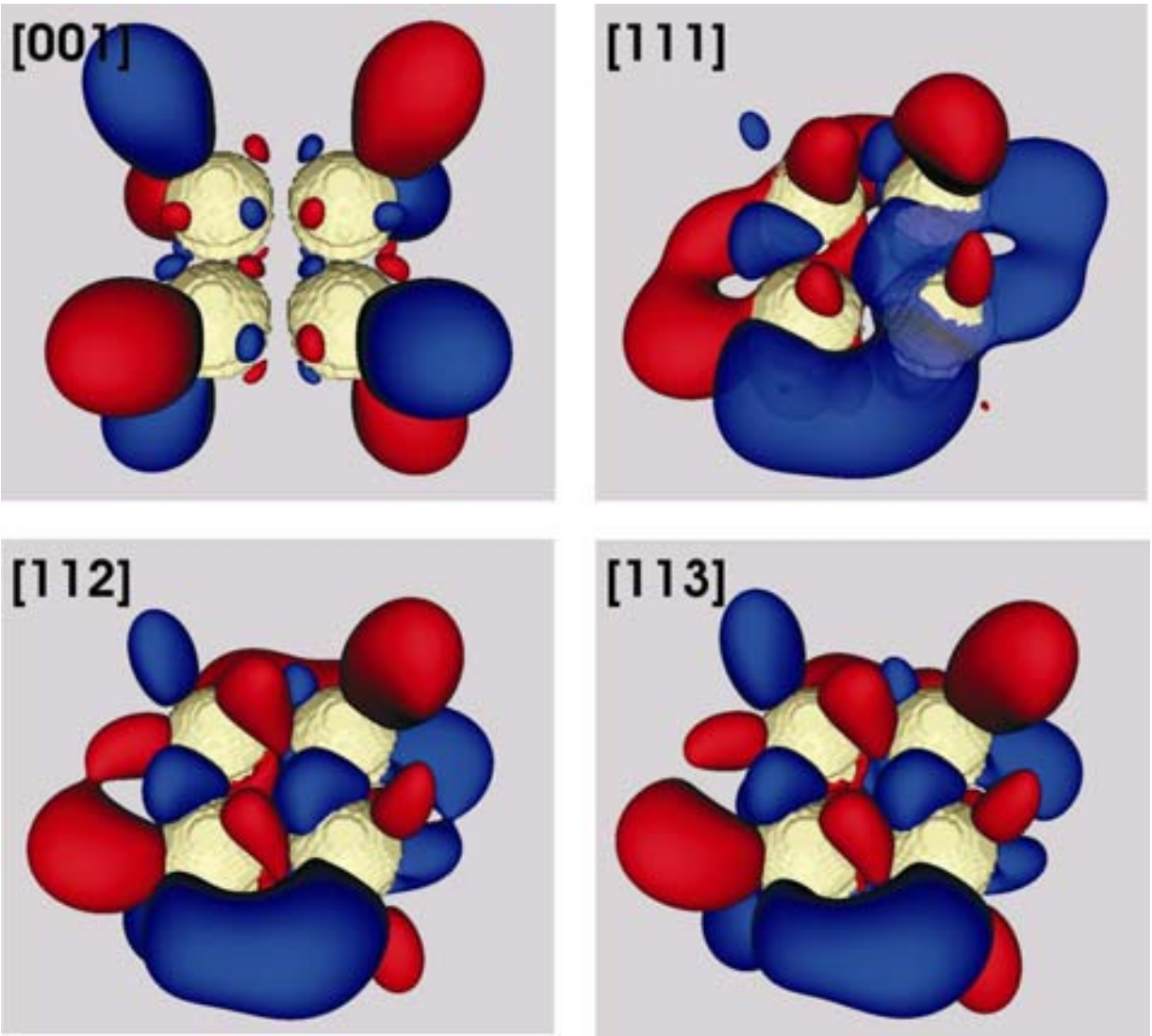

**Fig. 5 (Colour online)** Piezoelectric potential of M1 QDM with isosurfaces at ±32mV for different substrate orientation (blue -32mV, red +32mV).

Electron and hole confined states are extracted from the eight-band Hamiltonian including strain and piezoelectricity, and are shown in Fig. 6 for our model QDM. As we have already shown for the single QD case (see Ref. 17), electron states are mainly influenced by hydrostatic strain, while for hole states, there is a competition between strain effects and band mixing (increased for growth on high index surfaces). Depending on the size of the dot in the growth direction two regimes can be distinguished: (i) flat dots, where the variation of the hole energy levels with the substrate orientation is mainly influenced by the hydrostatic strain, and (ii) large dots, where band mixing, resulting from the kinetic part of the Hamiltonian, and reduced biaxial part of the strain tensor, has a dominant influence on the variation of the hole levels with substrate orientation. For the QDM the situation is much more complex. The electron ground state is mainly determined by the hydrostatic component of the strain tensor, whereas the electron ground state of the [111] grown QDM is always the highest in energy, and the one of the [113] grown QDM is the closest energetically to the reference case of [001] grown QDM. The qualitative behavior of the electron ground state in QDM is not influenced neither by the distance between the dots in the vertical direction, $d_{ver}$, nor by the inter-dot distances in the lateral direction, $d_{lat}$. The excited electron states, however, strongly depend on the variation of the inter-dot distances in the vertical direction.

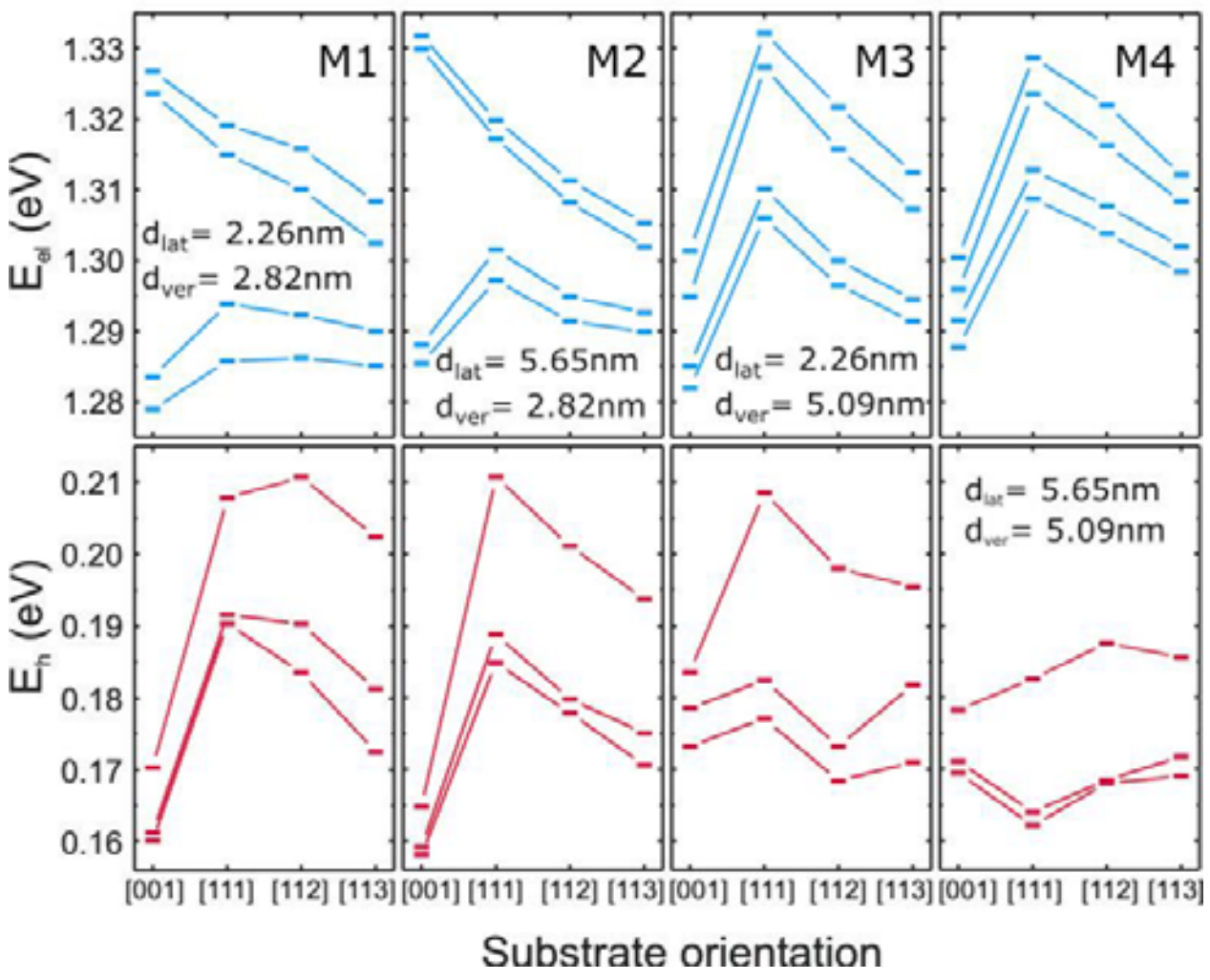

**Fig. 6 (Colour online)** Electron (upper panels) and hole (lower panels) energy levels as they vary with the substrate orientation for QDM. The electron and hole energy levels are given with respect to the top of the valence band of GaAs.The distance between the dots in lateral and growth direction is varied as well.

In contrast, the main influence on the hole states is coming from the kinetic part of the Hamiltonian and the biaxial component of the strain tensor supressing the influence of the hydrostatic component of the strain tensor. Increasing the inter-dot distances in the lateral direction, from M1 to M2 QDM, the lowest hole ground state energy changes from the one of [112] grown M1 QDM to [111] grown M2 QDM. The increase of the inter-dot distances in vertical direction, from M1 QDM to M3 QDM, has the same impact on the variation of the hole ground state energy with substrate orientation. However, the increase of the inter-dot distances in lateral and vertical direction, from M1 QDM to M4 QDM, the lowest energy has the hole ground state for [112] growth and is not altered by the change of inter-dot distances.

Transition energies of QDM, including direct Coulomb interaction in our calculations, as they vary with the substrate orientation are shown in Fig. 7. Depending on the distances between the dots forming the molecule, variation of the tranistion energies with the substrate orientation can be rather small, as is the case of the M3 molecule, or can qualitatively change their behaviour with the variation of M1 and M2 against M4. For example, for [111] grown molecules changing the inter-dot distance varies the transition energies up to 50meV. Therefore, our predictions reported here, can be used as a guideline for potential aplication of such a system, in e.g. optoelectronics. Note that a recent experiment (see Ref. 12) suggests the successful manipulation of the interdot distances *both* in the lateral and in the growth (i.e. vertical) direction.

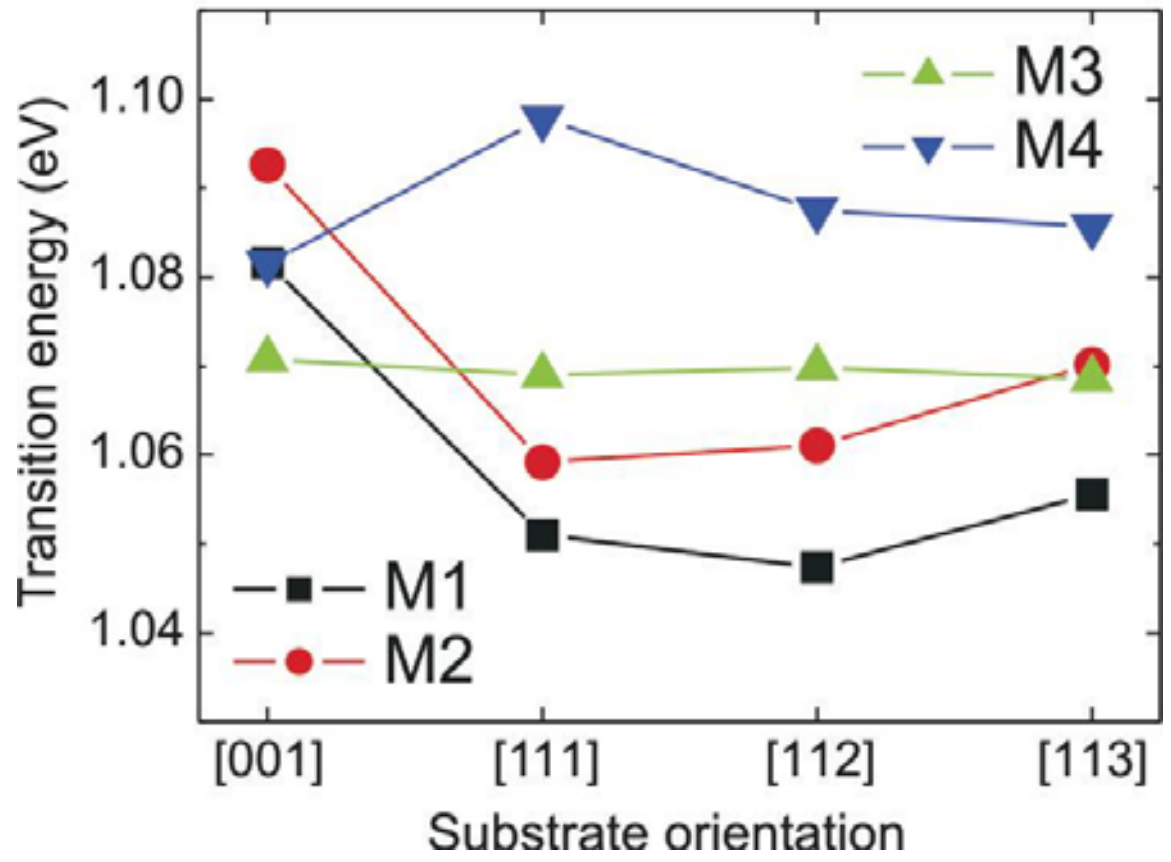

**Fig. 7 (Colour online)** Transition energies as they vary with the substrate orientation for QDMs.

## B. Effects of an external tilted magnetic field

Correct implementation of the magnetic field in the eight-band **k.p** model discretized on a grid is very important.

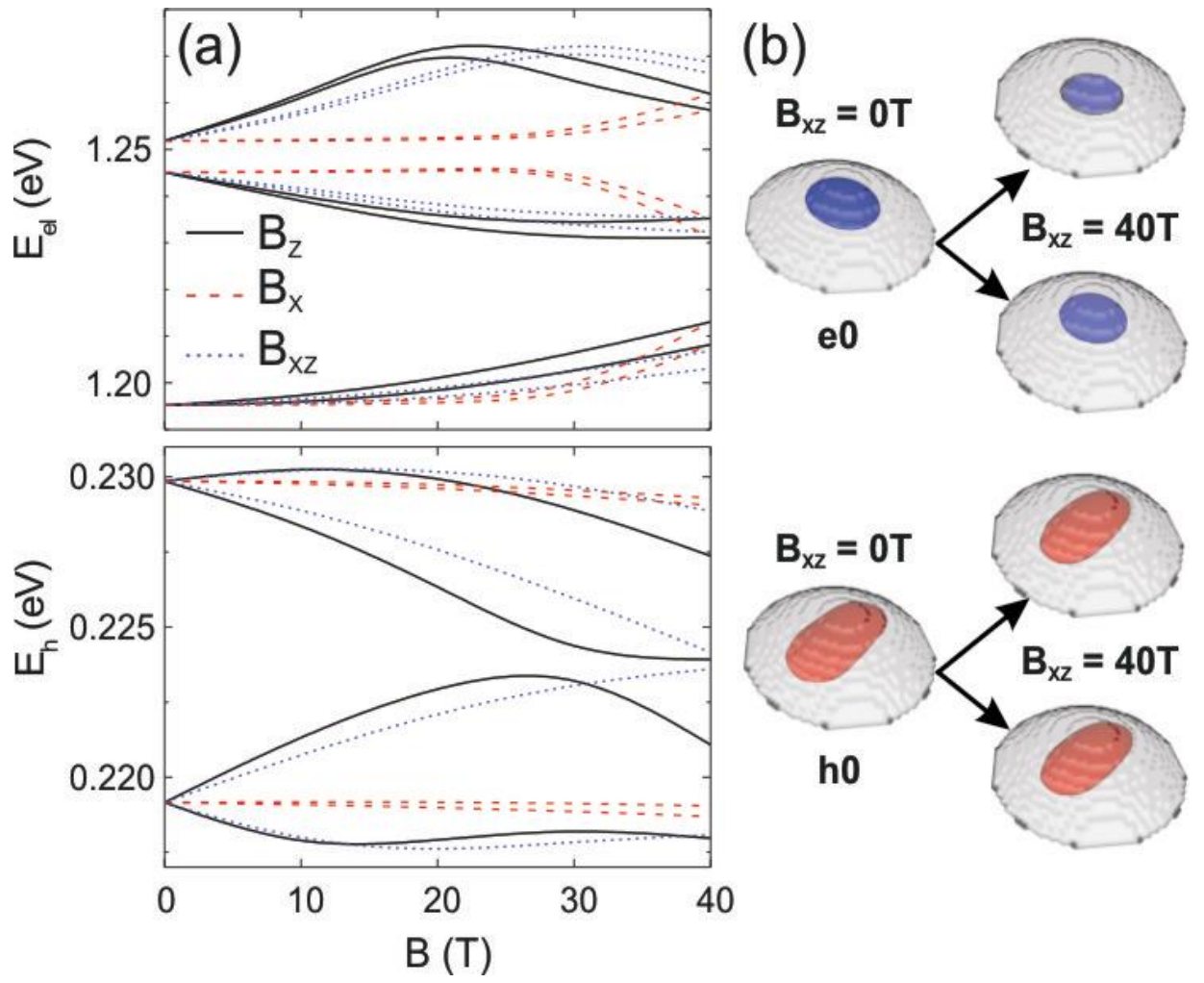

**Fig. 8 (Colour online)** (a) Electron (upper panel) and hole (lower panel) energy levels as they vary with magnetic field for [001] grown lens shaped QD. Three different directions of the magnetic field are shown: magnetic field parallel to the growth direction (z-axis) black solid curve, magnetic field in lateral direction (x-axis) – red dashed line, and magnetic field applied in the direction *π/4* between the *x*, and *z* axis – blue dotted lines (b) The square modulus of the electron and hole ground state wave functions for B = 0T, and $B_{xz}$ = 40T.

For symmetric structures in an external magnetic field, where both, electrons and holes are localized in the dot, implemetantion of a simple discretization scheme, i.e. by simply implementing the Peielers substitution in the wave vector and setting the origin of the coordinate system at the center of symmetry of the structure, would not lead to a qualitatively different (incorrect) behaviour of the eigenenergies and eignenvectors with variation of the magnetic field. Although the results so obtained are based on an inconsistent formalism, the results turn out to agree rather well with those obtained through a correct interpretation. As an example, we show the electron and hole states of the single [001] grown lens shaped QD of radius R = 13.56nm, and height h = 5.65nm placed in an external magnetic field, where the magnetic field is varied in the range 0T – 40T, and is directional dependent (see Fig. 8). If the origin of the coordinate system is placed in center of the dot, the difference in electron and hole energies and wave functions between two implementations is negligible. The competition between the quantum confinement and effects of an external magnetic field on the electron and hole energy levels of the lens shaped QD are shown in Fig. 8(a). The largest influence on the electron and hole energy levels is found for a magnetic field applied in the growth directions (the weakest lateral confinement). In Fig. 8(b) we show the electron and hole square modulus of the wave fuction at B = 0T and B = 40T, where the role of the magnetic field is reflected through the additional confinement and Zeeman splitting. Variation of the origin of the coordinate system, however, changes the eigenenergies and leads to numerically caused asymmetric wave functions. The problem also arises in the treatement of magnetic field effects of asymmetric structures, or e. g. type II QDs (where hole or electron are localized outside the dot) or QDM (see e.g. Ref.19).

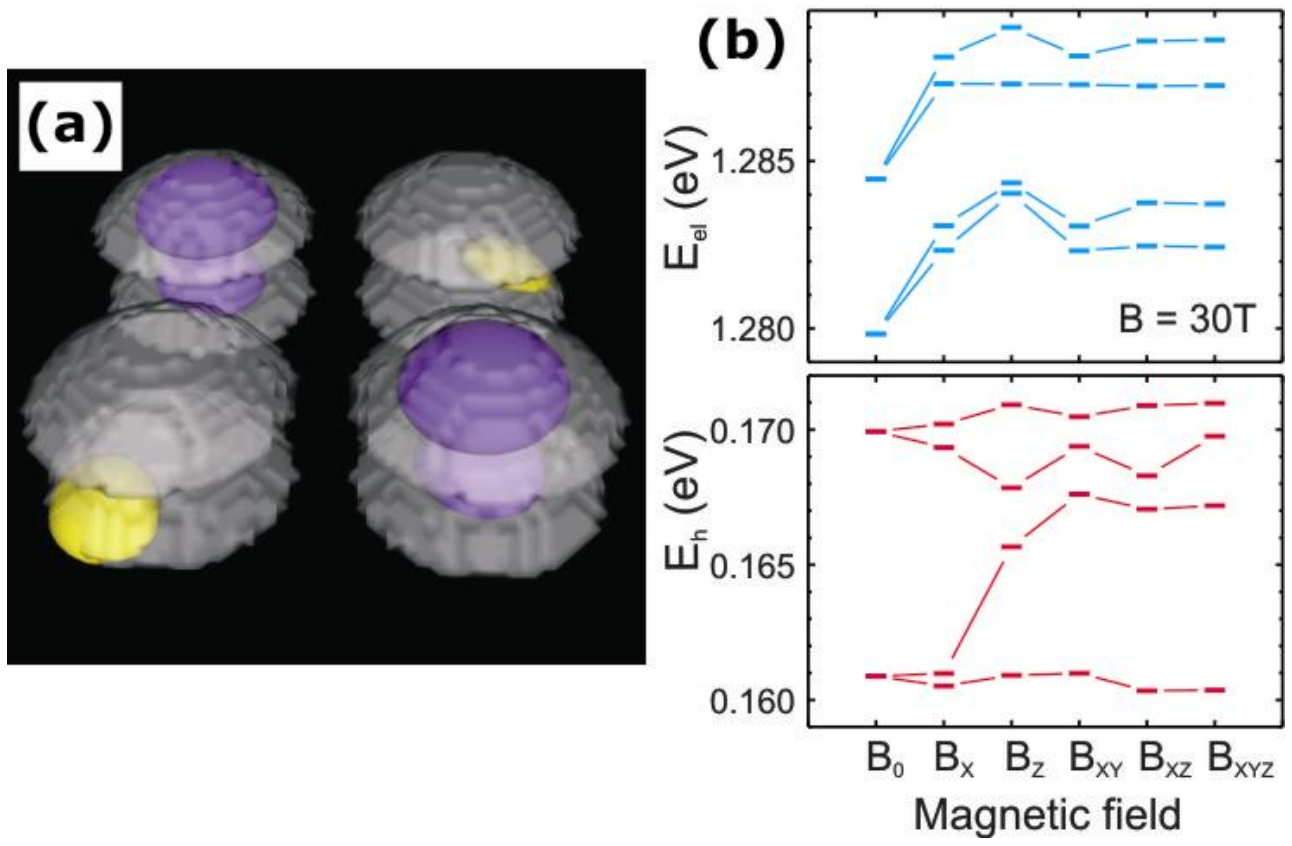

**Fig. 9 (Colour online)** (a) Probability density of the electron (violet) and hole (yellow) ground state in [001] grown M1 QDM. (b) Electron and hole energy levels for B=30T as they vary with the magnetic field direction. Subscript zero refers to zero magnetic field ($B_0$=0T), and subscripts *x*, *z*, *xy*, *xz*, *xyz* denote direction of magnetic field: *xy* -means that the magnetic field is applied in the direction *π/4* between *x*, and *y* axis. Other subscripts are defined in a similar way.

As an additional example, we consider here the [001] grown QDM consisting of eight identical lens shaped QDs in a magnetic field of 30T, where the direction of the magnetic field is varied (see Fig. 9). In our case the electron and hole are localized in different dots, as shown in Fig. 9(a) which is a direct consequence of the strain distribution and piezoelectric effect. Therefore, for such a system, the setting of the origin of the coordinate system at the center of the structure would lead to incorrect results. As an illustration, we show the electron and hole energy levels of QDM for B= 30T as they vary with the direction of the applied magnetic field. The results are shown in Fig. 9(b). The largest splitting of the electron ground state level in a magnetic field is found for the magnetic field applied in the direction π/4 between *x* and *z* direction, whereas the largest splitting for the hole ground state level is found to be for the case of magnetic field applied in the growth direction.

## IV.Conclusions

We presented a full 3D model for electronic structure calculations of QDs and QDMs grown on high index surfaces and/or in the presence of an external magnetic field. The strain distribution is calculated in the framework of continuum elasticity theory, and the single particle states are extracted from the nonsymmetrized eight-band Hamiltonian. Our model properly takes into account the effect of the substrate orientation by rotation of the coordinate system in the way that one coordinate coincides with the growth direction. The effects of a tilted external magnetic field are taken into account throught the Zeeman effect and employing a gauge invariant scheme based on Wilson's formulation of the lattice gauge theory.

We find that the piezoelectricity for QDs grown on [11*k*], where *k* = 1,2,3,4,5,7,9 and QDMs containing eight QDs grown on [11*l*], where *l* = 1,2,3 is much stronger and asymmetric as compared to the case of [001] grown dots and molecules, respectively. The [111] grown QDs and QDMs exibit the largest piezoelectric potential. We predicted the variation of the transition energies of QDM as a function of substrate orientation and interdot distances in the 8 QD molecule showing that the variation of the interdot distance qualitative changes the transition energy dependence on the substrate orientation. For example, for [111] grown molecules, changing the interdot distance varies the transition energies up to 50meV. We also addressed the question of the magnetic field direction dependent variation of the electronic properties of InAs/GaAs QDMs.

## Acknowledgements

This work was supported by the Belgian Science Policy (IUAP), the ESF–network AQDJJ, and the European Union Network of Excellence SANDiE.

## Notes and references